\documentclass[aps,prl,showpacs,twocolumn,superscriptaddress,bibnotes]{revtex4}

\usepackage{graphicx}
\begin{document}

\newcommand{\Hpip}{(C$_5$H$_{12}$N)$_2$CuBr$_4$}
\newcommand{\Hpips}{(Hpip)$_2$CuBr$_4$}
\newcommand{\Dpip}{(C$_5$D$_{12}$N)$_2$CuBr$_4$}
\newcommand{\Dpips}{(Dpip)$_2$CuBr$_4$}

\title{Direct Observation of Magnon Fractionalization in the Quantum 
Spin Ladder}

\author{B.~Thielemann}
\affiliation{Laboratory for Neutron Scattering, ETH Zurich and Paul Scherrer 
Institute, CH--5232 Villigen, Switzerland}

\author{Ch.~R\"uegg}
\affiliation{London Centre for Nanotechnology, University College London, London WC1E 6BT, United Kingdom}

\author{H. M.~R{\o}nnow}
\affiliation{Laboratory for Quantum Magnetism, Ecole Polytechnique 
F\'ed\'erale de Lausanne, CH--1015 Lausanne, Switzerland}

\author{A. M.~L\"auchli}
\affiliation{Max Planck Institut f\"ur Physik komplexer Systeme, 
N\"othnitzerstr. 38, D--01187 Dresden, Germany}

\author{J.--S.~Caux}
\affiliation{Institute for Theoretical Physics, University of Amsterdam,
1018 XE Amsterdam, The Netherlands}

\author{B. Normand}
\affiliation{Theoretische Physik, ETH--H\"onggerberg, CH--8093 Z\"urich, 
Switzerland }

\author{D.~Biner}
\author{K. W. Kr\"amer}
\author{H.--U.~G\"udel}
\affiliation{Department for Chemistry and Biochemistry, University of Bern, 
CH--3000 Bern 9, Switzerland}

\author{J.~Stahn}
\affiliation{Laboratory for Neutron Scattering, ETH Zurich and Paul Scherrer 
Institute, CH--5232 Villigen, Switzerland}

\author{K.~Habicht}
\author{K.~Kiefer}
\affiliation{BENSC, Helmholtz Centre Berlin for Materials and Energy, 
D--14109 Berlin, Germany}

\author{M.~Boehm}
\affiliation{Institut Laue Langevin, 6 rue Jules Horowitz BP156, 38024 
Grenoble CEDEX 9, France}

\author{D. F.~McMorrow}
\affiliation{London Centre for Nanotechnology and Department of Physics and 
Astronomy, University College London, London WC1E 6BT, United Kingdom}

\author{J.~Mesot}
\affiliation{Laboratory for Neutron Scattering, ETH Zurich and Paul Scherrer 
Institute, CH--5232 Villigen, Switzerland}
\affiliation{Laboratory for Quantum Magnetism, Ecole Polytechnique 
F\'ed\'erale de Lausanne, CH--1015 Lausanne, Switzerland}

\date{\today}

\begin{abstract}
We measure by inelastic neutron scattering the spin excitation spectra as 
a function of applied magnetic field in the quantum spin--ladder material 
\Hpip. Discrete magnon modes at low fields in the quantum disordered phase 
and at high fields in the saturated phase contrast sharply with a spinon 
continuum at intermediate fields characteristic of the Luttinger--liquid 
phase. By tuning the magnetic field, we drive the fractionalization of 
magnons into spinons and, in this deconfined regime, observe both 
commensurate and incommensurate continua. 
\end{abstract}


\maketitle

Gapped quantum antiferromagnets (AFs) offer some of the most exotic states 
of electronic matter ever to be observed, and as such have been the enduring 
focus of intense theoretical and experimental interest. In an applied 
magnetic field, the threefold degeneracy of the triplet excitations is 
lifted and the triplet gap $\Delta$ is closed at a critical field $B_c = 
\Delta/g \mu_{\rm B}$. The properties of the states beyond this quantum 
critical point (QCP) depend strongly on system dimensionality 
\cite{BEC_review}, and in one dimension (1D), where long--range order is 
forbidden, the field--induced phase is expected to be a spin Luttinger 
liquid (LL). In this state, spin--flip excitations fractionalize into 
spinons, elementary $S = 1/2$ entities, whose excitation spectrum is 
dramatically different from that of both ordered and truly quantum 
disordered (QD) magnets.

Here we present the results of inelastic neutron scattering (INS) 
measurements of the magnetic excitation spectrum of a quantum spin ladder. 
By tuning the magnetic field, we access the full ladder phase diagram, 
which includes QD, LL, and fully saturated (FM) phases. INS is the only 
experimental technique which can measure the full 
momentum-- and energy--dependence of the dynamical 
susceptibility, information which is essential to identify unambiguously 
the nature of the quasiparticles in these different phases, especially in 
the LL regime. We will show that both the QD phase (ladder magnetization 
$m = 0$) and the FM phase ($m = 1$) have well--defined magnon modes. In 
sharp contrast, the spectrum in the gapless LL state ($0 < m < 1$) is a 
continuum, whose spectral weight and incommensurate wave vector we control 
systematically.

\begin{figure}[t]
\begin{center}
\includegraphics[width=0.45\textwidth,clip]{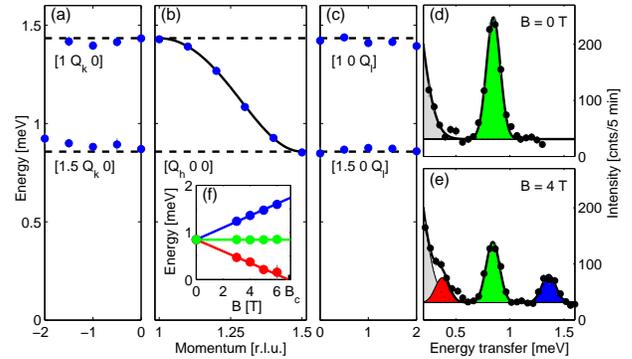}
\end{center}
\caption{\small Spin dynamics in the QD phase ($B < B_c$) at $T = 50$ mK. 
(a--c) Triplet dispersion at $B = 0$ from constant--$Q$ scans along the 
ladder axis and in the two perpendicular directions. (d--e) Triplet 
excitations at ${\mbox{\boldmath$Q$\unboldmath}} = [1.5\textrm{ }0\textrm{ 
}0]$ in fields of 0 and 4 T. (f) Zeeman splitting of the triplet modes 
at ${\mbox{\boldmath$Q$\unboldmath}} = [1.5\textrm{ }0\textrm{ }0]$. Solid 
lines are fits explained in the text.}
\label{fig1}
\end{figure}

Previous attempts to observe the essential physics of the field--induced LL phase in 
quasi--1D magnets such as the candidate Haldane material NDMAP \cite{rzc1}, 
effective $S = 1$ chain IPA--CuCl$_3$ \cite{rzc2}, and possible spin--ladder 
system CuHpCl \cite{B_CuHpCl} have generally encountered problems due to 
additional terms in the spin Hamiltonian. These include single--ion 
anisotropy, Dzyaloshinskii--Moriya interactions, and especially interchain 
couplings. While a spinon continuum has been measured in the gapless chain 
materials Cu--benzoate \cite{Dender}, CuPzN \cite{Stone}, and KCuF$_3$ 
\cite{Lake}, the spin--ladder compound piperidinium copper bromide 
[\Hpip]~\cite{Patyal,Watson} offers the first opportunity to induce 
LL physics in a gapped system by an applied magnetic field. A number of 
thermodynamic measurements, specifically of thermal expansion \cite{Lorenz}, 
specific heat \cite{rcrct}, and magnetocaloric effect \cite{rcrct,rbt3d}, 
as well as by nuclear magnetic resonance \cite{B_NMR}, are consistent with 
predictions for an ideal ladder. By INS the rung and leg exchange parameters, 
$J_r$ and $J_l$, as well as additional terms in the spin Hamiltonian, are 
determined directly.

High--quality single crystals of \Dpip~were grown from solution, 
and up to 10 were coaligned to obtain samples with masses of 
approximately 2.5 g. INS experiments were performed on the spectrometers 
IN14 (ILL, Grenoble), FLEX (HMI, Berlin), and TASP (SINQ, Villigen), using 
a focussing monochromator/analyzer and a Be filter between sample and 
analyzer (fixed final energies $E_f = 3.5$ meV or $E_f = 4.7$ meV). 
Cryomagnets were used for vertical fields up to 14.8 T ($B||b$--axis), 
and dilution inserts for temperatures down to 50 mK.

INS data in the QD phase, $B < B_c$, are summarized in Fig.~\ref{fig1}. 
Here we focus on the 1D ladder dispersion: sharp (resolution--limited) 
peaks arise from a dispersive triplet excitation for momentum transfers
along the ladder [$Q_h$, Fig.~\ref{fig1}(b)]. The dispersion in the 
perpendicular directions [$Q_k$ and $Q_l$, Figs.~\ref{fig1}(a,c)] is 
shown to demonstrate the excellent one--dimensionality of the system. 
In fact we have found a systematic variation of order $30 \mu{\rm eV}$ 
in the effective ladder bandwidth \cite{rbt3dins}, and subtract this 
interladder contribution to obtain the intrinsic parameters $J_r$ and $J_l$. 
Figures 1(d,e) show the INS intensity at ${\mbox{\boldmath$Q$\unboldmath}}
 = [1.5\textrm{ } 0\textrm{ } 0]$ (the band minimum), demonstrating a Zeeman 
splitting into three triplet components at finite field; Gaussian fits yield 
the energies shown in Fig.~\ref{fig1}(f).

\begin{figure}[t!]
\begin{center}
\includegraphics[width=0.45\textwidth,clip]{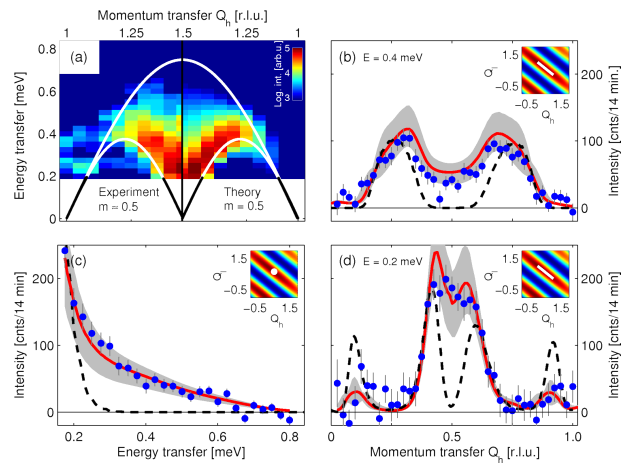}
\end{center}
\caption{\small Excitation spectrum in the LL phase ($T_{\rm N} < T$
 = 250 mK $< T_{\rm LL}$) at $B = 10.1$ T ($m \approx 0.5$) after 
subtraction of the zero--field background. (a) Measured (left) and 
simulated (right) INS intensities. Solid lines mark the edges of the 
two--spinon continuum. (b,d) Constant--$E$ scans taken along maxima 
of the transverse structure factor (insets: scan trajectories in 
white). (c) Constant--$Q$ scan at ${\mbox{\boldmath$Q$\unboldmath}} = 
[0.5\textrm{ } 0\textrm{ }  0.61]$. Black dashed lines in (b)--(d) are 
based on a $\delta$--function spinon spectrum [solid lines in panel (a)], 
red solid lines and shading on a full continuum calculation.}
\label{fig2}
\end{figure}

The spectrum changes dramatically above $B_c$: we find a continuum of 
excitations extending over much of the Brillouin zone and up to energies 
of 0.8 meV. We consider first a field corresponding to half magnetic 
saturation ($m = 0.5$, Fig.~\ref{fig2}), where the ladder is equivalent to 
a gapless spin chain in zero field (below). Here the continuous spectrum 
of spinon excitations \cite{Faddeev81} is bounded by $\epsilon_l (Q_h)
 = \hbar \omega (Q_h)$ and $\epsilon_u (Q_h) = 2 
\hbar \omega (\frac{Q_h}{2})$, where $\hbar \omega (Q_h) = \alpha 
J |\sin{(2 \pi Q_h)}|$ \cite{desCloizeaux62,Mueller81} with $\alpha$ 
a quantum renormalization factor which is determined exactly from the 
system geometry and interaction parameters. In Fig.~\ref{fig2}(a), it 
is clear that the commensurate $m = 0.5$ spectrum is well described 
by such a shape. Its continuum nature is illustrated strikingly in 
high--statistics measurements taken at $B = 10.1$ T: both constant--$E$ 
[Figs.~\ref{fig2}(b) and (d)] and constant--$Q$ [Fig.~\ref{fig2}(c)] scans 
show broad regions of continuous intensity, the latter extending from 0.15 
meV (lower measurement limit) to 0.8 meV. We stress two important points. 
First, these data are taken well inside the LL regime, at a temperature 
significantly below the LL crossover, $T_{\rm LL}$ \cite{rcrct}, but above 
the boundary to 3D order induced by interladder coupling ($T_{\rm N,max}
 = 110$ mK, \cite{B_NMR,rbt3d}). Second, the spinon continuum studied in 
Figs.~2 and 3 arises only from fractionalization of the lowest triplet 
branch of the QD phase.

\begin{figure*}[t!]
\includegraphics[width=0.75\textwidth,clip]{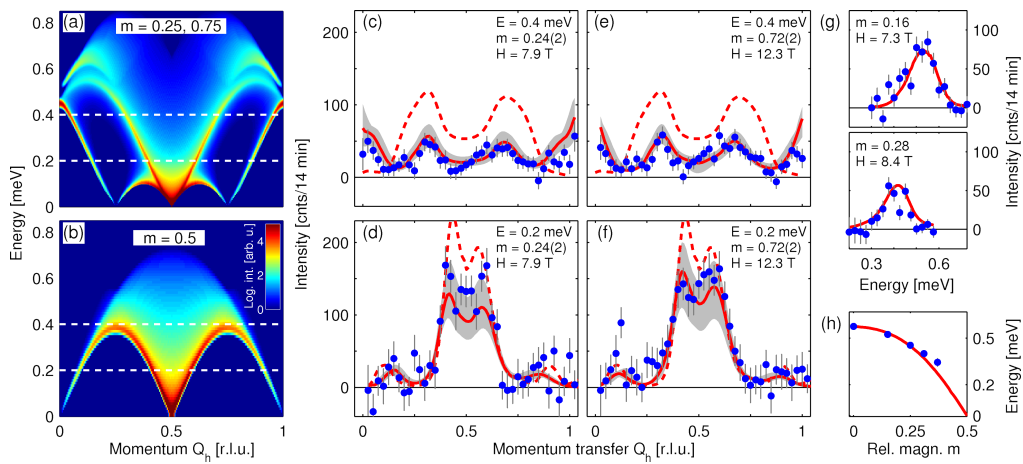}
\caption{\small Incommensurate excitations in the LL phase. 
(a,b) Spinon continua calculated for an $S = 1/2$ XXZ chain with 
anisotropy $\delta = 0.5$ (see text) at $m = 0.25, 0.75$ (a) and $m = 0.5$
[(b), data shown also in Fig.~2(a) convolved with instrumental resolution]; 
dashed white lines represent INS scans at $E = 0.2$ and $0.4$ meV. (c)--(f) 
Constant--$E$ scans (trajectories as in Fig.~2(b) inset) for $B = 7.9$ T
and $B = 12.3$ T. (g) Constant--$Q$ scans at the ZB for two chosen
magnetizations $m$ between 0 and 0.5. (h) Summary of data for ZB
excitation energy as a function of $m$. Red lines are predictions based
on the XXZ chain: solid from panel (a) and dashed (shown for comparison)
from panel (b).} 
\label{fig3}
\end{figure*}

We observe a continuous evolution of the excitation spectrum as the 
magnetic field is tuned away from $m = 0.5$. Figures \ref{fig3}(c)--(f) 
show constant--$E$ scans at $B = 7.9$ T and $B = 12.3$ T (corresponding 
approximately to $m = 0.25$ and $m = 0.75$) which are essentially identical 
within the experimental error. At $E = 0.4$ meV, the decrease of magnetic 
intensity compared to Fig.~2(b) between $Q_h = 0.2$ and $0.8$ is accompanied 
by additional weight around the zone boundaries (ZBs, which we define as 
$Q_h = 0$ and 1). For these values of $m$, field--induced shifts in spectral 
weight are less pronounced at $E =$ 0.2 meV. Because it is not possible by 
INS to follow the location of the zero--energy incommensurate point, instead 
we have measured the magnetic signal at the ZB as a function of field 
[Figs.~\ref{fig3}(g,h)], finding an increase in energy and intensity 
from $m = 0.5$ to $m = 0$ and 1.

\begin{figure}[t!]
\begin{center}
\includegraphics[width=0.47\textwidth,clip]{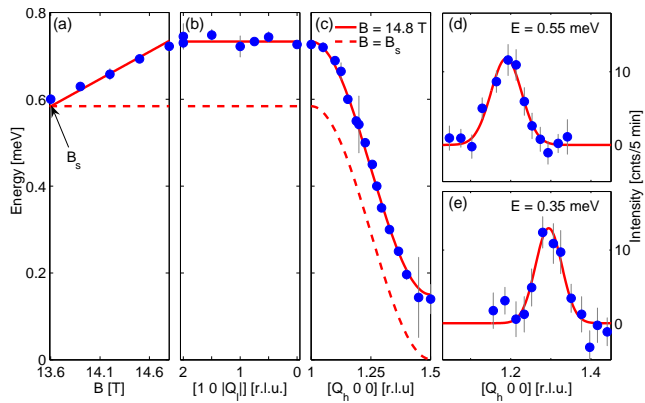}
\end{center}
\caption{\small Spin dynamics in the FM phase. (a) $B$--dependence of 
the ZB excitation energy. (b,c) Magnon dispersions along $Q_l$ and $Q_h$. 
(d,e) Constant--$E$ scans at $B = 14.8$ T and $T = 50$ mK with Gaussian 
fits. Solid lines in panels (a--c) are fits to the dispersion, while dashed 
lines allow extraction of $B_s$.  }
\label{fig4}
\end{figure}

When the magnetic field is increased beyond a second QCP at $B_s$, the spins 
are fully aligned. The spectrum becomes discrete again, as shown in 
Figs.~\ref{fig4}(d,e), with elementary magnon excitations. We observe a 
1D dispersive band [Figs.~\ref{fig4}(b,c)], whose width is very similar to 
that measured in the QD phase. The ZB excitation energy increases linearly 
with applied field [Fig.~\ref{fig4}(a)], which allows the identification of 
$B_s$.

We use the magnon dispersion relations to deduce the exchange 
parameters of the system. The excitations of the spin ladder in zero field 
have been the focus of much theoretical investigation \cite{Barnes93}. 
Because \Hpip~is rather ''strongly coupled'' ($J_r/J_l \approx 4$), 
high--order perturbative expansions are very effective, and here we follow 
the (3D) treatment of Ref.~\cite{rmm}. The measured triplet dispersion is 
dominated by the ladder terms: we obtain $J_r = 12.8(1)$ K and $J_l = 
3.2(1)$ K [black line in Fig.~\ref{fig1}(b)]. From the fits to the linear 
Zeeman splitting [solid lines in Fig.~\ref{fig1}(f)], the $g$--factor for 
this orientation is $g = 2.17(3)$, while $B_c = 6.8(1)$ T. The two 
approaches agree perfectly, and are consistent with values of $g$ and 
$B_c$ determined by other techniques.

Theoretically, the excitations of a field--polarized ladder 
are gapped spin waves with dispersion relation $\epsilon (Q_h) = 
g \mu_{\rm B} (B - B_s) + J_l (1 + \cos(2 \pi Q_h))$ \cite{Normand2000}. 
A mean--field treatment is exact here because all quantum fluctuations are 
quenched. The red lines in Figs. \ref{fig4}(b,c) are fitted using this 
expression: when the small 3D coupling term is removed, we obtain $J_r = 
13.1(1)$ K, $J_l = 3.3(1)$ K, and $B_s = 13.6(1)$ T. The fit to the ZB 
energy in Fig.~\ref{fig4}(a) yields $B_s = 13.6(2)$ T. 

The LL continuum arises from the fractionalization of spin--flip 
excitations ($\Delta S^z = 1$) into two elementary and deconfined $S = 1/2$ 
objects. Indeed, in a unified description of the full phase diagram, the 
magnon excitations in the QD and FM phases are bound states of these 
spinons, and the QCP at $B_c$ may be regarded as a field--driven spinon 
binding--unbinding or confinement--deconfinement transition. The theoretical 
description for a ladder in the strong--coupling limit may be obtained by a 
mapping to the $S = 1/2$ XXZ chain with anisotropy $\delta = 1/2$ 
\cite{Mila98}. The low--energy sector of the ladder [{\it i.e.}~the 
lowest Zeeman--split branch in Fig.~1(f)] is governed by 
\begin{equation}
 \mathcal{H}_{\rm XXZ} = \sum_i J_l (\tilde S^x_i \tilde S^x_{i+1} + 
\tilde S^y_i \tilde S^y_{i+1} + \delta \tilde S^z_i \tilde S^z_{i+1})
 - b_{\rm eff} \tilde S^z_i, 
\label{ehxxz}
\end{equation}
where the effective field $b_{\rm eff} = 2 b_s \frac{B - (B_c + B_s)/2}{B_s
 - B_c}$ is such that $- b_s \le b_{\rm eff} \le b_s$, with $b_s = 
\frac{3}{2} J_l$ the saturation field for the XXZ chain. This effective 
model has spinon excitations in whose dispersion $J = J_l$ and $\alpha
 = 1.299$ \cite{Mueller81}, whence the bounds $\epsilon_l (Q_h)$ and 
$\epsilon_u (Q_h)$ shown as solid lines in Fig.~\ref{fig2}(a). Because 
${\mbox{\boldmath${\tilde S}$\unboldmath}}_i$ in Eq.~(\ref{ehxxz}) is a 
composite of the two physical spins on each ladder rung, the total INS 
cross--section contains a rung structure factor modulating the contributions 
from longitudinal and transverse spin correlations \cite{rbt3dins}. All data 
in the LL phase were measured on the maxima of the transverse structure 
factor (insets in Fig.~\ref{fig2}), where the longitudinal contributions 
are zero.

We have calculated the transverse spin correlation function for all 
values of $m$ following Ref.~\cite{Caux05}, and in Figs.~\ref{fig3}(a,b) 
present the results for $m = 0.25$, 0.75, and 0.5. In the XXZ--chain 
model, the spectral intensities are symmetric in $m$ about $m = 0.5$. Data 
taken at $B = 10.1$ T correspond to $m = 0.48(2)$, while fields $B = 7.9$ T 
and $B = 12.3$ T correspond respectively to $m = 0.24(2)$ and $m = 0.72(2)$. 
The theoretical intensities were convolved with the 4D instrumental 
resolution to obtain the global fit shown as red lines in 
Figs.~\ref{fig2}(b)--(d) and \ref{fig3}(c)--(f), while the red lines in 
Figs.~\ref{fig3}(g,h) are obtained directly. The shaded bands indicate 
the error bar in the experimental determination of a single constant of 
proportionality valid for all fields, energies, and wave vectors. Their
width combines the statistics of all our scans with uncertainties in the
exact magnetization values at the chosen fields and in the convolution 
procedure. The agreement is quantitatively excellent. We note in particular 
that only the smallest asymmetries between $m < 0.5$ and $m > 0.5$ may be 
discerned in the data [{\it cf.}~Figs.~3(c,d) and 3(e,f)], and that the 
energy--dependence of the intensity is described exactly [Fig.~2(c)].

Measurements in the QD and FM phases provide two independent and
complementary determinations of $J_r$ and $J_l$. While the values of 
$B_c$ and $B_s$ deduced from these are marginally smaller than from 
thermodynamic studies, the latter were performed mostly on undeuterated 
samples and in different orientations. By exploiting the 
${\mbox{\boldmath$Q$\unboldmath}}$--specificity of INS, we have accounted 
for a 3\% effect from interladder coupling \cite{rbt3dins}. We then find 
that the intrinsic ladder leg parameters in the two regimes are identical 
within their errors, but the rung parameters are not: this apparent 
magnetostriction effect is of order 1--2\%. That such a phenomenon
may occur is not surprising in a structurally "soft" material of this
nature \cite{Lorenz}. In fact this discrepancy is the sum of all 
additional contributions, including any other magnetoelastic terms 
or complex spin interactions. Our magnon dispersion analysis therefore 
quantifies the statement that \Hpip~is an excellent spin--ladder system.

A key property of the LL spectrum is the presence of a zero--energy mode 
at an incommensurate wave vector $0 < Q_{\rm min} < 1/2$ which changes 
systematically with field. However, the spectral weight at $Q_{\rm min}$ 
vanishes as $E \rightarrow 0$ [Fig.~3(a)], precluding a direct measurement 
of the incommensurability. Instead we have presented indirect confirmation 
of the theoretical prediction in the form of the field--tuned finite--$E$ 
spectra and the ZB energy [Figs.~2 and 3]. The question of the evolution 
of spectral weight in the LL is of particular interest in the context of 
the commensurate 3D ordered phase which emerges at sufficiently low 
temperature \cite{B_NMR,rbt3d}. At intermediate energies, we return to the 
question of the symmetry of the measured spectra about $m = 0.5$: physical 
effects arising due to departures from strict strong coupling, and from the 
higher triplet branches, are expected to cause some asymmetry in intensities, 
but these are clearly extremely small (for energies $0.5 J_l < E < 2.5 J_l$) 
in \Hpip. At high energies, it remains to address, both experimentally and 
theoretically, the nature of the higher spinon continua expected from the 
upper two triplet branches.

In summary, we have performed a comprehensive INS investigation of the 
magnetic excitation spectrum in \Hpip, a spin--ladder compound whose energy
scales are perfectly suited to systematic studies in laboratory fields. We 
observe the presence of a broad continuum of spinon excitations in the 
intermediate, Luttinger--liquid (LL) phase, which is starkly different from 
the discrete (magnon) excitations measured below the critical field (QD phase) 
and above saturation (FM phase). From the QD and FM results, we extract the 
ladder parameters with unprecedented accuracy, demonstrating directly that 
even the sum of all other effects beyond the "nearly ideal" Hamiltonian of 
weakly--coupled ladders falls below the 2\% level. The spinon continuum 
proves the occurrence of field--induced fractionalization into elementary 
$S = 1/2$ entities as the system enters the LL regime. An excellent, fully 
quantitative description of the incommensurate continua measured at all 
fields is obtained from an effective chain model for a ladder with the 
coupling ratio $J_r/J_l \approx 4$ of \Hpip.

This project was supported by the Swiss National Science Foundation  
(Division II and the NCCR MaNEP), the Royal Society, EPSRC and FOM. This work 
is based partially on experiments performed at the Swiss spallation neutron 
source, SINQ, at the Paul Scherrer Institute.

\end{document}